# Pre-examinations Improve Automated Metastases Detection on Cranial MRI

Deike-Hofmann, Katerina MD[1, 2]; Dancs, Dorottya MD[1]; Paech, Daniel MD[1]; Schlemmer, Heinz-Peter MD[1]; Maier-Hein, Klaus PhD[3]; Bäumer, Philipp MD[1]; Radbruch, Alexander MD, JD[2]; Götz, Michael PhD[3]

[1]Department of Radiology, German Cancer Research Center, Heidelberg
[2]Department of Neuroradiology, Bonn University Clinic, Bonn
[3]Department for Medical Image Computing, German Cancer Research Center, Heidelberg, Germany.

## Abbreviations

Artificial Intelligence, AI; Computer Aided Detection, CAD; Change Detection, CD; Machine Learning, ML; melanoma brain metastases, MM; false positive, FP; false negative, FN; true positive, TP; positive predictive value, PPV; cranial MRI, cMRI

## ABSTRACT

OBJECTIVE

To assess the diagnostic value of inclusion of pre-diagnosis MRI and different MRI sequences when training a convolutional neural network (CNN) in detection of metastases from malignant melanoma (MM) on an annotated real-life cranial MRI dataset. Diagnostic performance was challenged by extracerebral-intracranial MM and by inclusion of MRI with varying sequence parameters.

MATERIALS AND METHODS

Our local ethics committee approved this retrospective monocenter study.

Firstly, a dual time approach was assessed, for which the CNN was provided sequences of the MRI that firstly depicted new MM (diagnosis MRI) as well as of a pre-diagnosis MRI:
Inclusion of only contrast-enhanced T1-weightings ($CNN_{dual_ce}$) was compared to inclusion of also the native T1-weightings, T2-weightings and FLAIR sequences of both time points ($CNN_{dual_all}$). Secondly, results were compared to the corresponding single time approaches, in which the CNN was provided exclusively the respective sequences of the diagnosis MRI. Case-wise diagnostic performance parameters were calculated from five-fold cross validation.

RESULTS

In total, 94 cases with 494 MM were included. Overall, the highest diagnostic performance was achieved by inclusion of only the contrast-enhanced T1-weightings of the diagnosis and of a pre-diagnosis MRI ($CNN_{dual_ce}$, sensitivity = 73 %, PPV = 25 %, F1-score = 36 %). Using exclusively contrast-enhanced T1-weightings as input resulted in significantly less false positives (FPs) compared to inclusion of further sequences beyond contrast-enhanced T1weightings (FPs = 5 / 7 for $CNN_{dual_ce}$ / $CNN_{dual_all}$, $p < 1e\text{-}5$). Comparison of contrastenhanced dual and mono time approaches revealed that exclusion of pre-diagnosis MRI significantly increased FPs (FPs = 5 / 10 for $CNN_{dual_ce}$ / $CNN_{ce}$, $p < 1e\text{-}9$).
Approaches with only native sequences were clearly inferior to CNNs that were provided contrast-enhanced sequences.

CONCLUSIONS

Automated MM detection on contrast-enhanced T1-weightings performed with high sensitivity. Frequent FPs due to artifacts and vessels were significantly reduced by additional inclusion of pre-diagnosis MRI, but not by inclusion of further sequences beyond contrastenhanced T1-weightings. Future studies might investigate different change detection architectures for computer-aided detection.

## **INTRODUCTION**

Radiologic imaging is of ever-growing importance in patient care and the number of acquired scans per patient increases exponentially - as does the radiologists´ working load. This especially holds true for oncologic imaging, wherefore this study investigated computeraided detection (CAD) of metastases from malignant melanoma, the malignancy with the fastest growing incidence[1], on cranial MRI. Fortunately, the introduction of immunotherapies increased overall survival of patients with advanced melanoma[2–4], who are screened from head to thigh every three months accumulating approximately 3.000 images per follow-up.

Therefore, intelligent algorithms for image interpretation have a large market and are welcomed broadly[5–9]. Several studies investigated CAD systems for automated assessment of cerebral pathologies[10–19]. However, the realization of CAD systems is highly variable depending on the issue at question, machine learning (ML) architecture and input data. Therefore, this study aimed to investigate the influence of different input data on performance of a convolutional neural network (CNN). CNNs are a family of algorithms especially suited for image analysis, and have also been applied in different contexts for the comparison of image pairs[20–22].

In human reading the availability of different MRI sequences increases specificity as a lesion has to meet a specific signal pattern to be classified as a potential MM. Features increasing the probability for presence of MM include verifiability on different sequences, contrastenhancement, hyperintensity on native T1-weighting, susceptibility artifacts and diffusion restriction. Therefore, we hypothesized that CAD performance benefits from a multi sequence input that allows extraction of the multifarious MM imaging features.
Besides sequence plurality, the assessment of lesion dynamics over time is of considerable importance to classify unclear findings as either malignant or benign, wherefore we investigated whether CAD performance profits from inclusion of pre-diagnosis MRI, i.e. whether change detection is superior to single time approaches.

In summary, this study assessed the diagnostic performance of a CNN trained in MM detection dependent on different sequence input.

## PATIENTS AND METHODS

### Patients and melanoma brain metastases

Written informed consent was waived due to the retrospective character of this IRB approved study. A total of 494 MM on 94 diagnosis MRI of 43 patients with histologically proven malignant melanoma were included. Mean age (range) of the patients was 69 (44 - 94) years, comprising 8 women and 35 men. Total number of included time points was 115, with a mean number of 2.7 (2.0 – 6.0) per patient and a mean time interval of 161 (range 29 – 958, standard deviation (SD) ± 154) days between two examinations. Median number (25% - 75% quantile) of MM per case (i.e. an MRI pair or single diagnosis MRI for the dual / mono time approach, respectively) was 3.0 (2.0-4.0) and 7.0 (2.0-6.5) per patient. Median (25% - 75% quantile) lesion diameter was 4.2 mm (3.0 mm – 7.1 mm) and lesion volume was 58 $mm^3$ (25 $mm^3$ - 232 $mm^3$).
The dataset comprised the following metastases that would have been not or worse detectable if brain extraction would have been performed: 16 dural as well as 8 leptomeningeal metastases located on the brain convexity, furthermore one subcutaneous as well as one osseous metastasis. Reliable detection of this MM can only be trained by inclusion of the entire cMRI without brain extraction.

### MRI protocol

No exclusion criteria due to varying MRI parameters, differing field strengths, vendors or contrast agents were applied. The majority of MRI scans was performed at our institution (86.2 %, 81/94) at a 1.5 Tesla MRI system (67.9 %, 55/81, for sequence parameters see Table, Supplemental Digital Content 1, which illustrates the Siemens Magnetom Symphony 1.5 Tesla MR protocol) using a standard dose of Gadoterate meglumine (Dotarem®, 0.1 mmol/kg body weight).
Only patients with availability of the following MRI sequences at the time of initial MM diagnosis (hereinafter referred to as diagnosis MRI) as well as on a preceding MRI (prediagnosis MRI) were included: Native as well as contrast-enhanced T1-weighting (T1w and ceT1w, respectively), T2-weighting (T2w) and fluid-attenuated inversion recovery MRI (FLAIR).

### Dataset preparation

Entire MRI data processing was conducted in-house. The four input sequences (T1w, ceT1w, T2w and FLAIR) were exported to the Medical Imaging Interaction Toolkit (MITK v2018.04.2). Metastases annotation was performed with the free-hand tool on the ceT1w of the diagnosis MRI, which was sampled to an 0.5x0.5mm in plane resolution. Each MM was annotated on all slices, on which it was detectable. The sequences of the same time point were registered and resampled to the ceT1w image space. Two time points were registered by mapping the ceT1w sequence of the pre-diagnosis MRI to the diagnosis MRI and applying the found transformation to all other sequences of the pre-diagnosis MRI. For all registration and resampling steps, an automatic affine registration with mattes mutual information as metric and a regular step gradient descent as optimizer followed by 3D-BSpline interpolation from the MatchPoints MultiModal.affine.default module within MITK Phenotyping (v2018-1018) was applied[23,24].

MM annotation was performed by one of two readers after a teaching session with a radiologist with 9 years of experience in neuroimaging. Unclear lesions were discussed and classified in consensus. Identification of MM was performed on an accredited workstation with access to the complete clinical sequence protocol. MRI reports and medical histories of the patients were taken into account.

## Machine learning architecture

The here used CNN is a modification of the so-called U-Net, a CNN that was developed for biomedical image segmentation, basing on the fully convolutional network[25] and modified to work with fewer training images and to yield more precise segmentations.

During the contracting path of the U-Net, which is a typical convolutional network that consists of repeated application of convolutions, spatial information is decreased and feature information is increased. The subsequent expansive pathway combines the feature and spatial information through a sequence of up-convolutions and concatenations with highresolution features from the contracting path.

The architecture of the here used CNN was modified in accordance to Daudt et al. and is presented in Figure 1A[26].

## Experiment

Diagnostic accuracy of the just introduced CNN was assessed with respect to varying input data. Initially, dual time approaches, i.e. CNNs trained with both the diagnosis MRI as well as a pre-diagnosis MRI, were compared with respect to different input sequences. Initially, the performance of a CNN which was provided all four MRI sequences (i.e. T1w, T2w, FLAIR and ceT1w) of the diagnosis and pre-diagnosis MRI ($CNN_{dual_all}$) was compared to the performance resulting from training exclusively with the ceT1w of both time points ($CNN_{dual_ce}$).

Subsequently, diagnostic performance parameters of the combination of all native sequences combined ($CNN_{dual_native}$) as well as separately ($CNN_{dual_T1}$, $CNN_{dual_T2}$, $CNN_{dual_FLAIR}$) and the combination of the ceT1w of the diagnosis MRI and the native T1w of the pre-diagnosis MRI were assessed. To investigate the additional benefit of inclusion of the FLAIR sequence, a CNN was also trained with both ceT1w and FLAIR of both time points ($CNN_{dual_FLAIR_ce}$).

Secondly, the performance of the contrast-enhanced CNNs was compared to their mono time equivalents, for which only the diagnosis MRI, but not the pre-diagnosis MRI, was provided ($CNN_{all}$, $CNN_{ce}$ and $CNN_{FLAIR_ce}$).

## Training

Training was performed from scratch, i.e. without any sort of pre-training from other datasets. An in-house implementation of the described U-Net in Python (version 3.7.6) using PyTorch (version 1.4) as implementation backbone for the deep learning part was used for the experiments.

Due to five-fold cross validation, 80 / 20 % of the total dataset served as training / test set, respectively. In turn, 75 / 25 % from the current training fold served as training / validation set. The network was trained over 70 epochs without any early stopping criterion using a learning rate of 0.0002 and a batch size of 4. SoftDice together with an additional equally weighted CrossEntropy term as regularizer[27] were used as loss in combination with the Adam optimizer. The image dimensions were set to 512x512x3. Image augmentation was performed by mirroring along all three axis as well as variating contrast and noise with the corresponding augmentation methods using the default parameters from Batchgenerators (version 0.17). All hyperparameters were chosen based on experience and past experiments without optimization on the test folds.
The experiments ran on our internal cluster and local workstations with varying hardware configurations. The minimum hardware used for these experiments was a Nvidia TitanX with 12 GB of memory and 32 GB of working memory. Training of a CNN took less than half a day, with the actual run time depending on the hardware and the number of sequences. The generation of the prediction mask of a single slice took 0.4 seconds on average, the one of the entire cranial MRI less than 30 seconds. This might prospectively allow for automatic calculation of the CNN output before radiological assessment[28].

## CNN output

The CNN output, i.e. the inferred segmentations of the CNNs, were smoothed by dilation with a range of six voxels in order to fuse solitary, tightly neighbored segmentations. This simulated human image interpretation, in which adjacent areas are simultaneously inspected. Each fully connected neighborhood counted as an individual lesion and a lesion counted as true positive (TP) if the connected neighborhoods from the manual or inferred mask showed an overlap of at least one voxel. Accordingly, lesions on the manual / inferred mask without at least one correlating voxel on the other mask counted as false negatives (FNs) or false positives (FPs), respectively. True negative lesions were not reported as negative lesions were not defined.

## Statistics

Case-wise TPs, FPs, FNs as well as sensitivity (TP / (TP + FN)), PPV (TP / (TP + FN)) and F1-score (2xTP / (2xTP + FN + FP)) were calculated from five-fold cross validation. We reported mean (±SD), median (25% and 75% quantiles) and 95 % confidence intervals (CI) for all parameters.
A case was defined as either an MRI-pair (diagnosis plus pre-diagnosis MRI) or exclusively the diagnosis MRI for the dual or mono time approach, respectively.
The F1-score is the harmonic mean of sensitivity and PPV and is commonly used for evaluating detection tasks. It corresponds to the Dice Coefficient when used for evaluating segmentations[29].

Paired Wilcoxon test was used to test for significant different performance parameters of the CNNs. As 25 significance tests were performed, a Bonferroni corrected significance level of $p_{Bonferroni} < 0.002$ (highly significant: $p_{Bonferroni} < 0.0004$) was applied.

## RESULTS

A total of 494 MM on 94 diagnosis MRI were included. All test results that undercut the significance level ($p_{Bonferroni} < 0.002$) were actually found to be also highly significant ($p_{Bonferroni} < 0.004$).

Table 1 presents TPs, FNs and FPs, while Table 2 indicates median sensitivity, PPV and F1score with 25% and 75% quantiles, and 95%-CIs of the different CNNs. Because performance parameters do naturally not show Gaussian distribution due to their prescribed range, the median was reported. However, mean and SD showed high accordance (compare Table, Supplemental Digital Content 2, which provides corresponding means and standard deviations of sensitivity, PPV and F1-scores dependent on sequence input).

Figure 2 presents contrast-enhanced MRI scans of a representative case at time of diagnosis of new MM as well as the corresponding output of the $CNN_{dual_ce}$.

Figure 3 compares sensitivity, PPV and F1-scores of the different CNNs.

### Evaluation of dual time approaches

**All vs. contrast-enhanced sequences:** A median sensitivity of 72.7% (CI [66.7% - 91.4%]) was achieved by inclusion of only the ceT1-weightings of the diagnosis and pre-diagnosis MRI ($CNN_{dual_ce}$). When trained on all four sequences of both time points ($CNN_{dual_all}$), an inferior median sensitivity of 50.0% (CI [50.0% - 71.4%]) was attained, however, this difference was not significant compared to the Bonferroni-adjusted significance level (p-value = 4.9e-2). Furthermore, number of FPs was significantly smaller for the $CNN_{dual_ce}$ (5, CI [4 – 6]) compared to the $CNN_{dual_all}$ (7, CI [5 – 9], p-value < 1e-5). Consecutively, PPV and F1score raised significantly from 15.4% (CI [11.1% - 23.5%]) to 25.0% (CI [23.0% - 33.3%]) and 25.0% (CI [18.2% - 28.6%]) to 36.4% ([28.6% - 42.4%]) by exclusion of sequences beyond contrast-enhanced T1-weighting (p-values < 1e-3), respectively.

**Native sequences**: The best performance of CNNs trained exclusively with native sequences ($CNN_{dual_native}$, $CNN_{dual_nT1}$, $CNN_{dual_T2}$, $CNN_{dual_FLAIR}$) was achieved by the $CNN_{dual_FLAIR}$ with a sensitivity, specificity and F1-score of 16.7%, 9.1%, and 14.3%, respectively. Therewith, the $CNN_{dual_FLAIR}$ was significantly inferior to the $CNN_{dual_ce}$ (p-values < 1e-6) and no further assessment of native CNNs was performed.

### Evaluation of mono time approaches

To investigate, how further reduction of input sequences impacts diagnostic performance, contrast-enhanced mono time approaches were assessed and compared to their mono time equivalents. Mono time approaches included exclusively sequences of the diagnosis MRI as CNN input; sequences of the pre-diagnosis MRIs were not provided.

**Performance of mono time approaches:** There was no significant difference between both contrast-enhanced mono time approaches ($CNN_{all}$ and $CNN_{ce}$) in terms of sensitivity, PPV and F1-score.

**Comparison of mono time approaches with their dual time equivalents**: With high median sensitivities greater than 70% for the mono time approaches $CNN_{all}$ and $CNN_{ce}$, sensitivities did not differ significantly between the contrast-enhanced mono and dual time approaches. However, the exclusion of pre-diagnosis MRIs increased FPs for the allsequence and the contrast-enhanced-only approach from 7 (CI [5-9]) to 10 (CI [9-11]), and 5 (CI [4-6]) to 10 (CI [8-12]), respectively. This led to higher PPVs and F1-scores of the dual time approaches, however only the difference in PPV and F1-score between $CNN_{ce}$ and $CNN_{dual_ce}$ was significant (p-values < 1e-4), not the ones between $CNN_{all}$ and $CNN_{dual_all}$.

In summary, the experiments showed that the dual time ceT1w-based approach ($CNN_{dual_ce}$) significantly outperformed all other approaches in terms of F1-score and performed at least similar with regards to sensitivity and PPV.

**Evaluation of further dual time approaches**

To investigate whether additional sequence combinations could further improve the diagnostic performance of the CNN, the following input combinations were evaluated and compared to the best solution found in the previous experiments (i.e. $CNN_{dual_ce}$).

**Evaluation of inclusion of the native T1-weighting (nT1w) of the pre-diagnosis MRI:** To investigate, whether inclusion of the pre-diagnosis native T1-weighting adds diagnostic benefit to automatic metastases detection, a CNN was provided the ceT1w of the diagnosis MRI as well as the nT1w of the pre-diagnosis MRI ($CNN_{T1n_ce}$). Sensitivity, PPV and F1-score did not differ significantly between $CNN_{T1n_ce}$ and $CNN_{ce}$, wherefore it was concluded that inclusion of the pre-diagnosis nT1w did not add diagnostic benefit to CNN performance.

**Evaluation of addition of the Flair sequence to the ceT1w:** While the $CNN_{dual_ce+FLAIR}$ did not differ significantly from the $CNN_{dual_ce}$ with regards to sensitivity, PPV and F1-score, the $CNN_{ce+FLAIR}$ was significantly inferior to the $CNN_{dual_ce}$ in terms of PPV and F1-score (p-values < 1e-6).

## DISCUSSION

This study is the first to assess the diagnostic benefit of inclusion of pre-diagnosis MRI as well as of the different clinical MRI sequences for automated detection of cranial melanoma metastases by a convolutional neural network. In contrast to previous studies on lesion assessment on cranial MRI, we did not perform brain extraction and challenged algorithm performance by inclusion of meningeal, intraosseous and subcutaneous metastases from malignant melanoma[10,12–15,17]. Furthermore, an uncensored real-life dataset was applied admitting non-uniformity of MRI systems and sequence parameters, which further complicated learning of MM imaging features.

Against the just mentioned challenges and despite the small size of the MM (median diameter of 4.2mm), our study revealed a high sensitivity for MM detection when the contrast-enhanced T1-weighting was included. Furthermore, false positives were significantly reduced by providing a pre-diagnosis contrast-enhanced scan, but not by inclusion of further sequences, which rather lowered specificity.

Although contrast-enhancement is known to be an imaging hallmark of brain metastases, the specific imaging features of MM, which result from their potential melanin content, can increase specificity of the human reader, which we erroneously hypothesized to apply equally for reading by AI. While one might argue that a greater training set might have made the CNN learn the more complex imaging pattern of MM including possible outliers, the here found dominant role of the contrast-enhanced scan concurs well with results from human readings which found that MM manifest initially with disruption of the blood-brain-barrier[30].

This study provides evidence that inclusion of pre-examinations significantly increased specificity. While reduction of FPs might have also been achieved by choosing a more strict operating point, i.e. raising the threshold of when a CNN output is counted as an MM, this was not carried out to avoid overfitting. Furthermore, we considered sensitivity more important than specificity in the clinical setting because FPs proposed to the radiologist can easily be denied while FNs might have severe therapeutical consequences.

The diagnostic benefit of pre-examinations for computer-aided detection is presumably caused by the same phenomenon as in human image interpretation: While solitary snapshots reveal a multitude of potential lesions - including a great number of FPs such as blood vessels and artifacts - new lesions or lesions with changed dimensions turned out to be more likely TPs than unchanged lesions.

Limitations have to be acknowledged. Firstly, the multi sequence approach did not include diffusion- (DWI) and susceptibility-weighted (SWI) MRI, which might have added value to ML performance. However, inferior resolution of DWI and SWI, if applicable at all, might have masked their potential benefit. Accordingly, two retrospective analyses did find neither susceptibility artifacts nor diffusion restriction to be early MRI findings of MM.[30,31]

Secondly, sensitivity of the CNNs was highly variable. This is likely caused by the great heterogeneity of the data, firstly, in terms of MM morphology such as size, occurrence of microbleedings, degree of contrast-agent enhancement and susceptibility artifacts and secondly due to varying MRI parameters, which differed inter- as well as intra-individually. On short notice, this heterogeneity surely resulted in a lower detection performance in our limited dataset, however, we deliberately didn´t homogenize the data but set great value on robustness and generalizability in the long run. While scalability is a major advantage of machine learning, this advantage can only be exploited when algorithms are unrestrictedly applicable due to training on authentic data that allows learning of robustness against common data flaws and artifacts and iteratively improves performance[32].

Last but not least, a radiologist reading of the entire clinical MRI protocol served as reference for MM classification and segmentation. Obviously, even though highly desirable, a histologically proven presence or absence of MM for every potential lesion is not possible due to ethical constraints. However, therapy decisions also base on MM imaging and not on histopathology. In future, a near-optimal ground-truth might be achieved by segmentations of multiple readers, unsupervised ML completely independent of any reference or training with practically infinite large datasets that allow the algorithm to abstract an optimal feature model of MM lesions that iteratively overcomes the less optimal model of an earlier learning stage.

Potential consequences for future clinical routine, derived from our study, include potential timesaving by computer-aided MM detection not just in terms of reading time of the radiologist, but also in terms of patient-scanner-time due to reduction of input sequences. In contrast, an inalienable value of contrast agent application was confirmed which is of considerable importance as many patients and clinicians are increasingly critical of serial gadolinium contrast agent application due to potential allergic reactions, renal impairment and gadolinium deposition[33–36].

The here used CNN concatenated the input sequences before passing them through the network, treating the first time point just like additional sequences of the second time point. Instead of the applied early fusion architecture, realization of change detection in deeper parts of the network might be investigated prospectively[37,38]. Furthermore, automatic assessment of more complex MM dynamics such as evaluation of therapy response might be addressed in future studies.

## FIGURE LEGENDS

**Figure 1**, Presentation of architecture, input and output of the applied convolutional neural network (CNN)

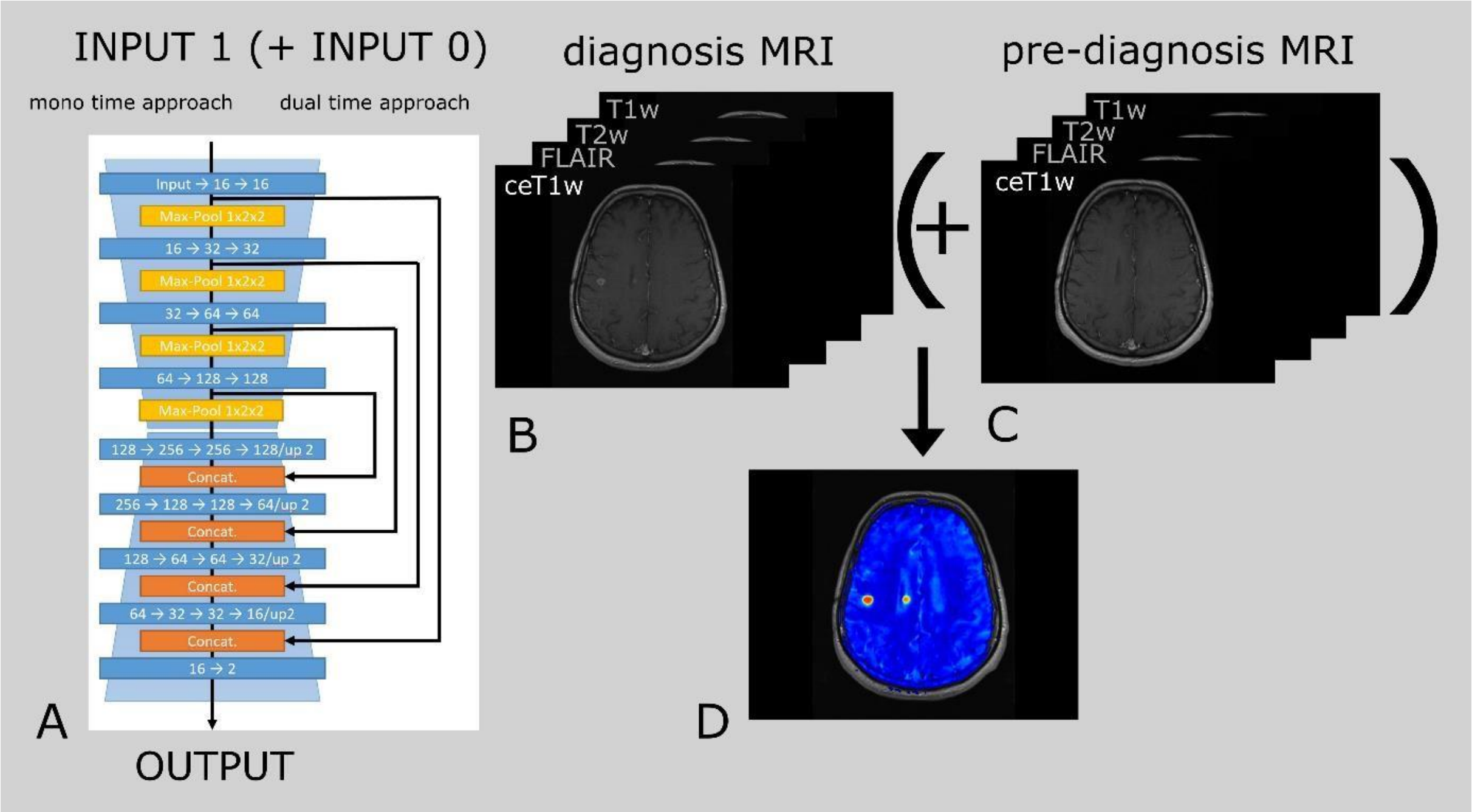

**(A)** The CNN architecture is based on the U-Net-architecture, which consists of a contracting and an expansive path (pictured by the light blue hourglass-shape). During contraction, the spatial information is reduced with a combination of convolutions (blue, numbers indicate feature channels) and max pooling (yellow) while feature information is increased. The expansive arm combines the feature and spatial information through a sequence of (up-) convolutions and concatenations (orange) with integration of high-resolution features from the contracting path (black arrows).

**(B)** Mono time approaches included exclusively sequences of the diagnosis MRI while **(C)** dual time approaches also comprised sequences of an MRI prior to initial diagnosis of melanoma metastases (MM). **(D)** The output of the CNN was a probability heat-map, i.e. an overlay of the input sequence highlighting areas of high MM probability (here: two true positive MM in the right hemisphere).

**Figure 2**, Example of results of automated melanoma brain metastases (MM) detection by a convolutional neural network (CNN) trained on contrast-enhanced T1-weightings (ceT1w) of the diagnosis and of a pre-diagnosis MRI ($CNN_{dual_ce}$)

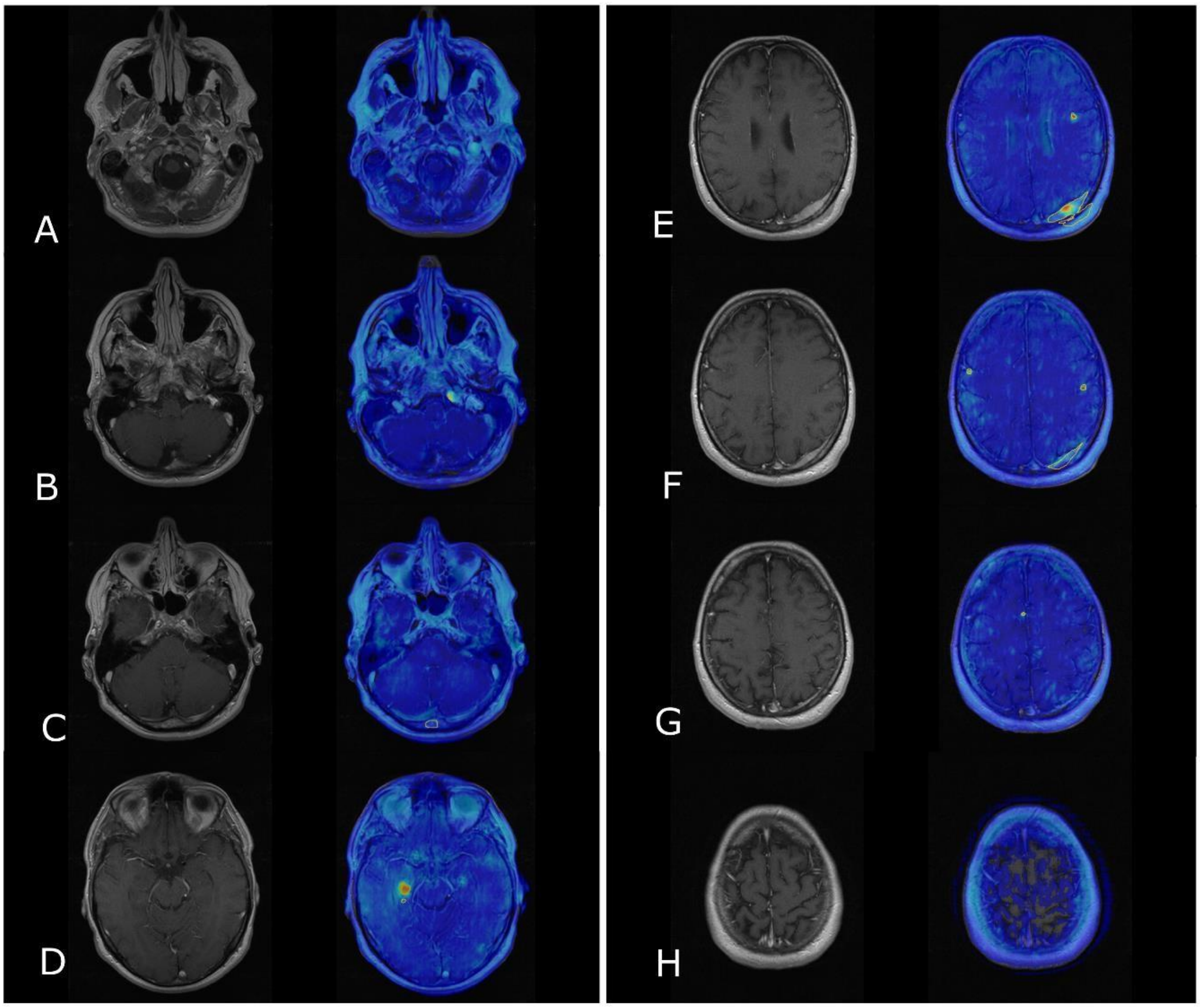

**A-E** Left column depicts ceT1w of the diagnosis MRI, right column presents the CNN output as a probability heat map overlay of the ceT1w. Yellow lines present MM annotation by a radiologist. All slices with any $CNN_{dual_ce}$ findings are presented; slices not shown were negative for true and false positives (TPs and FPs, respectively, compare **A** and **H**). Five of seven MM were correctly detected by the $CNN_{dual_ce}$ (compare **E-G**), including a large dural MM with intraosseous and subcutaneous infiltration (**E**).

Two MM were missed by the CNN, one of them located in the occipital bone (**C**) and the other one medially to the anterior horn of the right lateral ventricle (**D**).

Two FPs were recorded in the left para-pontine soft tissue (**B**) as well as in the right choroid plexus (**D**).

**Figure 3**, Comparison of sensitivity, positive predictive value (PPV) and F1-score dependent on input data

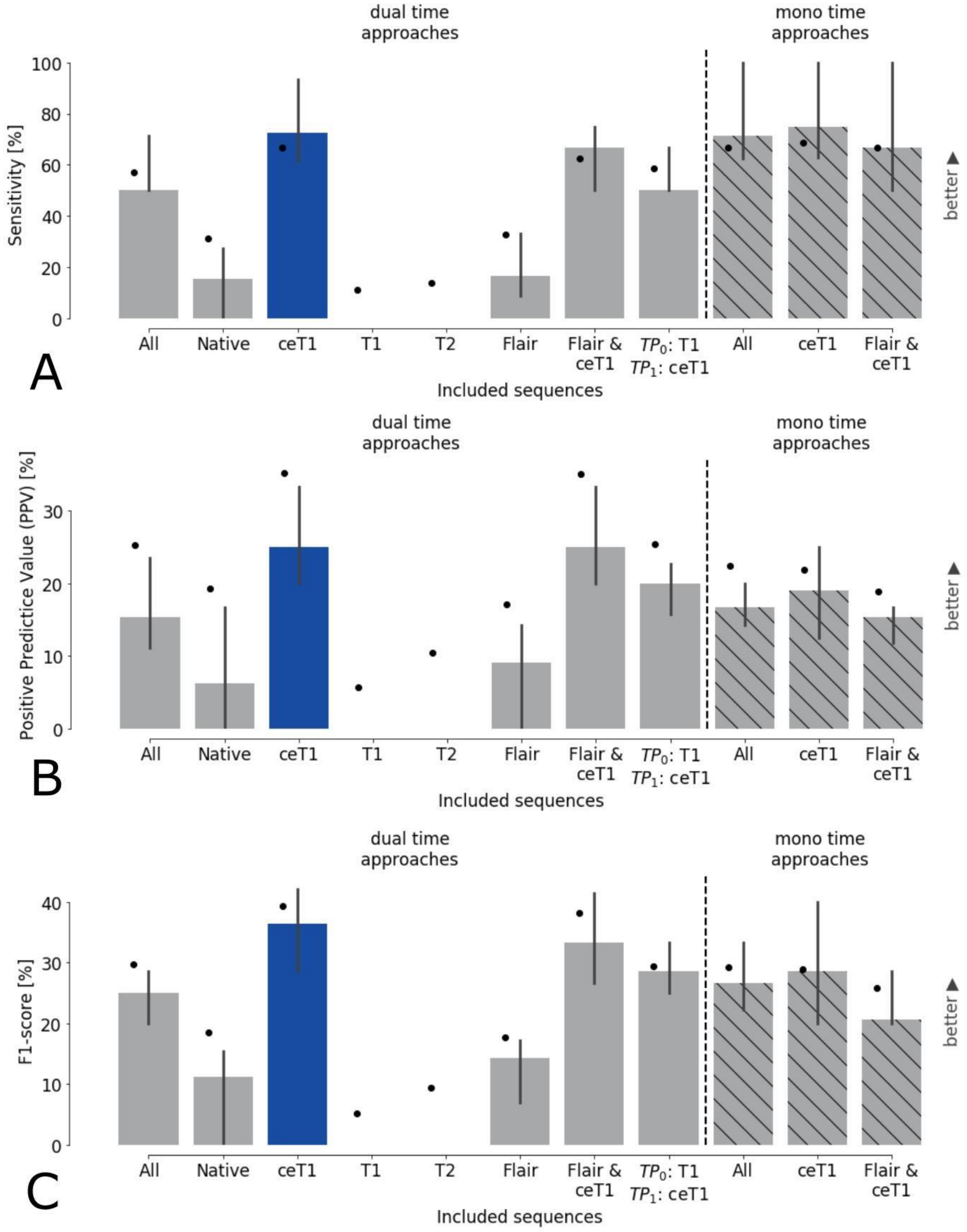

Depiction of median (column hight), 95% confidence interval of the median (CI, vertical lines) and mean (dots) with regards to the different input sequences.

From left to right: All, all four sequences served as CNN input (contrast-enhanced T1weighting (ceT1), native T1-weighting (T1), T2-weighting (T2) and fluid-attenuated inversion recovery MRI (Flair)). Native, all native sequences (T1w, T2w, FLAIR) were provided as sequence input.

The combinations of Flair with ceT1 (Flair & ceT1) as well as of the ceT1w of the diagnosis and the native T1 of the pre-diagnosis MRI were also assessed ($TP_0$: T1, $TP_1$: ceT1).
Non-hatched columns present the results of the dual time approach, in which the sequences of both the diagnosis MRI of melanoma brain metastases as well as a of a pre-diagnosis MRI served as algorithm input, while hatched columns depict the results of the mono time approach following provision of exclusively the diagnosis MRI.

A) No significant difference in median sensitivity was found between the approaches that included the ceT1w sequence. However, CNNs with only native sequences were clearly inferior in terms of sensitivity.

B) Positive predictive values (PPVs) of the dual time contrast-enhanced approaches (non-hatched columns) were higher compared to their mono time equivalents (hatched columns) due to reduction of false positives by inclusion of pre-diagnosis MRI.

C) Overall, F1-score of the dual time ceT1w approach (blue column) was highest.

**Table 1. Mean and median with confidence interval of true positives, false negatives and false positives of the different CNNs**

| CNN | Time point s inclu ded | Seque nce input | True Positives | | | False Negatives | | | False Positives | | |
|---|---|---|---|---|---|---|---|---|---|---|---|
| | | | Mean ± SD | Median (CI) | 25-75 % Quantils | Mean ± SD | Median (CI) | 25-75 % Quantils | Mean ± SD | Median (CI) | 25-75% Quantils |
| $CNN_{dual_all}$ | 1st diagnosis MRI AND prediagnosis MRI | ceT1w, T1w, T2w, FLAIR | 2.8 ± 4.4 | 1 (1-2) | 1-3 | 2.4 ± 4.8 | 1 (1-1) | 0-2 | 12.7 ± 15.4 | 7 (5-9) | 3-14 |
| $CNN_{dual_ce}$ | | ceT1w | 3.1 ± 4.9 | 2 (1-2) | 1-3 | 2.2 ± 4.7 | 1 (1-1) | 0-2 | 6.6 ± 6.6 | 5 (4-6) | 2-9 |
| $CNN_{dual_native}$ | | T1w, T2w, FLAIR | 1.3 ± 2.4 | 1 (0-1) | 0-1 | 3.9 ± 6.4 | 2 (1-2) | 1-4 | 4.8 ± 5.5 | 3 (2-5) | 2-6 |
| $CNN_{dual_nT1}$ | | T1w | 0.6 ± 1.9 | 0 (0-0) | 0-0.5 | 4.6 ± 7.2 | 2 (2-3) | 1-4 | 6.4 ± 6.1 | 5 (3-6) | 2-9 |
| $CNN_{dual_T2}$ | | T2w | 0.6 ± 1.2 | 0 (0-0) | 0-1 | 4.6 ± 7.6 | 2 (2-2) | 1-4 | 4.5 ± 3.6 | 4 (3-5) | 2-6 |
| $CNN_{dual_FLAIR}$ | | FLAIR | 1.2 ± 1.5 | 1 (1-1) | 0-2 | 4.0 ± 7.3 | 1 (1-2) | 1-4 | 6.8 ± 5.3 | 5 (4-7) | 3-9 |
| $CNN_{ce+FLAIR}$ | | ceT1w, FLAIR | 2.7 ± 3.5 | 1 (1-2) | 1-3 | 2.6 ±5.6 | 1 (1-1) | 0-1 | 6.9 ± 7.9 | 4 (3-5) | 2-8 |
| $CNN_{T1n_ce}$ | | TP 0: T1w, TP1: ceT1w | 2.6 ± 3.9 | 2 (1-2) | 1-2 | 2.6 ± 5.1 | 1 (1-1) | 0-2 | 10.5 ± 9.3 | 8 (6-10) | 4-15.5 |
| $CNN_{all}$ | 1st diagnosis MRI only | ceT1w, T1w, T2w, FLAIR | 3.0 ± 3.9 | 2 (1-2) | 1-3 | 2.3 ± 4.8 | 1 (0-1) | 0-2 | 11.1 ± 7.6 | 10 (9-1) | 6-13 |
| $CNN_{ce}$ | | ceT1w | 3.1 ± 4.9 | 2 (1-2) | 1-3 | 2.1 ± 4.7 | 1 (0-1) | 0-2 | 12.8 ± 12.1 | 10 (8-12) | 5-15 |
| $CNN_{ce+FLAIR}$ | | ceT1w, FLAIR | 2.7 ± 3.1 | 2 (1-2) | 1-3 | 2.5 ± 6.1 | 1 (1-1) | 0-2 | 13.2 ±9.8 | 10 (9-13) | 5-16.5 |

**Abbrev.:** CNN, convolutional neural network; ceT1w, contrast-enhanced T1-weighting; T1w, native T1-weighting; T2w, T2-weighting; FLAIR, fluid-attenuated inversion recovery MRI; SD, standard deviation; n, number; CI, 95% confidence interval of the median

**Table 2. Comparison of diagnostic performance parameters dependent on sequence input**

| **CNN** | **Time points included** | **Sequence input** | **Sensitivity [%]** | | **PPV [%]** | | **F1-score [%]** | |
|---|---|---|---|---|---|---|---|---|
| | | | **Median (CI)** | **25-75% Quantils** | **Median (CI)** | **25-75% Quantils** | **Median (CI)** | **25-75% Quantils** |
| $CNN_{dual_all}$ | diagnosis MRI AND prediagnosis MRI | ceT1w, nT1w, T2w, FLAIR | 50.0 (50.0-72.7) | 30.8-100 | 15.4 (11.1-23.5) | 5.7 - 36.7 | 25.0 (20.0-28.6) | 9.8 - 41.7 |
| $CNN_{dual_ce}$ | | ceT1w | 72.7 (66.7-93.3) | 50.0-100 | 25.0 (20.0-33.3) | 13.7 - 50 | 36.4 (28.6-42.4) | 19.1 - 57.1 |
| $CNN_{dual_native}$ | | nT1w, T2w, FLAIR | 15.4 (0-31.8) | 0.0-50.0 | 0.0 (16.733.3) | 0 - 33.3 | 11.1 (0-18.2) | 0 - 34.0 |
| $CNN_{dual_nT1}$ | | nT1w | 0 (0-0) | 0-4.5 | 0 (0-0) | 0 - 1.8 | 0 (0-0) | 0 - 3.1 |
| $CNN_{dual_T2}$ | | T2w | 0 (0-0) | 0-28.1 | 0 (0-0) | 0 - 13.5 | 0 (0-0) | 0 - 20.0 |
| $CNN_{dual_FLAIR}$ | | FLAIR | 16.7 (8.840.0) | 0-51.7 | 9.1 (3.7-13.3) | 0 - 22.3 | 14.3 (6.917.2) | 0 - 23.8 |
| $CNN_{ce+FLAIR}$ | | ceT1w, FLAIR | 66.7 (50.0-75) | 45.8-100 | 25.0 (20.0-33.3) | 13.2 - 50.0 | 33.3 (26.7 - 41.5) | 20.7 - 52.8 |
| $CNN_{T1n_ce}$ | | TP 0: T1w, TP1: ceT1w | 50.0 (50.0-66.7) | 44.7-100 | 20.0 (15.8 - 22.7) | 5.9 - 37.5 | 28.6 (25.0 - 33.3) | 9.5 - 40.7 |
| $CNN_{all}$ | diagnosis MRI only | ceT1w, nT1w, T2w, FLAIR | 71.4 (62.5-100) | 50.0-100 | 16.7 (14.3-20.0) | 7.1 - 30.8 | 26.7 (22.2-33.3) | 13.3 - 41.1 |
| $CNN_{ce}$ | | ceT1w | 75.0 (66.7-100) | 50.0-100 | 19.0 (12.5-25.0) | 7.4 - 30.0 | 28.6 (20.0-40.0) | 13.3 - 42.9 |
| $CNN_{ce+FLAIR}$ | | ceT1w, FLAIR | 66.7 (50.0 - 100) | 50.0-100 | 15.4 (12.5-16.7) | 7.1 - 29.5 | 20.7 (20.0 - 28.6) | 12.9 - 36.4 |

**Abbrev.:** PPV, positive predictive value; ceT1w, contrast-enhanced T1-weighting; FLAIR, fluid-attenuated inversion recovery MRI, n, number; CI, 95 % confidence interval

**List of Supplemental Digital Content**

**Supplemental Digital Content 1**. Table that illustrates the Siemens Magnetom Symphony

1.5 Tesla MR protocol. Pdf, page 1

**Supplemental Digital Content 2**. Table that illustrates mean and standard deviation of sensitivity, PPV and F1-score dependent on sequence input. Pdf, page 2

**Supplemental Digital Content 3**. Table that presents the performed significance tests with resulting p-values. Pdf, page 3

**Supplemental Digital Content 1.** Siemens Magnetom Symphony 1.5 Tesla MRI protocol

| | T1w | T2w | ceT1w | FLAIR |
|---|---|---|---|---|
| TE [ms] | 12 | 101 | 17 | 98 |
| TR [ms] | 500 | 5200 | 525 | 8000 |
| ST [mm] | 4 | 4 | 4 | 5 |
| B-value [s/mm²] | - | - | - | - |
| TI [ms] | - | - | - | 2340 |
| No. of Averages | 1 | 2 | 1 | 2 |
| Acq. Time [min:sec] | 4:00 | 3:53 | 4:00 | 5:38 |
| Field of View | 230 | 230 | 230 | 230 |
| No. of slices | 37 | 37 | 37 | 30 |
| Flip angle [°] | 90 | 150 | 90 | 150 |
| Acquisition Matrix | 0/320/234/0 | 0/320/205/0 | 0/320/234/0 | 0/320/205/0 |

**Abbr.:** T1w, native T1-weighted; T2w, T2-weighted; ceT1w, contrast-enhanced T1weighted; FLAIR, fluid-attenuated inversion recovery; TE, Echo Time; TR, Repetition
Time; TI, Inversion Time; ST, Slice Thickness; Acq. Time, Acquisition Time

**Supplemental Digital Content 2,** Mean and standard deviation of sensitivity, PPV and F1-score dependent on sequence input

| CNN | Time points | Sequence input | Sensitivity [%] | | PPV [%] | | F1-score [%] | |
|---|---|---|---|---|---|---|---|---|
| | | | mean | ±SD | mean | ±SD | mean | ±SD |
| $CNN_{dual_all}$ | diagnosis MRI AND prediagnosis MRI | ceT1w, nT1w, T2w, FLAIR | 57.1 | 35.4 | 25.2 | 26.1 | 29.7 | 25.4 |
| $CNN_{dual_ce}$ | | ceT1w, | 66.9 | 33.8 | 35.2 | 29.7 | 39.4 | 25.8 |
| $CNN_{dual_native}$ | | nT1w, T2w, FLAIR | 31.1 | 37.7 | 19.3 | 26.5 | 18.5 | 22.5 |
| $CNN_{dual_nT1}$ | | nT1w | 11.2 | 24.3 | 5.7 | 13.4 | 5.3 | 10.5 |
| $CNN_{dual_FLAIR}$ | | FLAIR | 32.6 | 22.7 | 10.4 | 20 | 9.5 | 14.6 |
| $CNN_{dual_T2}$ | | T2w | 13.9 | 35.6 | 17.2 | 23.3 | 17.7 | 21.3 |
| $CNN_{ce+FLAIR}$ | | ceT1w, FLAIR | 62.6 | 35.2 | 35.1 | 29.6 | 38.2 | 25.9 |
| $CNN_{T1n_ce}$ | | TP 0: T1w, TP1: ceT1w | 58.5 | 34.3 | 25.4 | 25.3 | 29.5 | 22.3 |
| $CNN_{all}$ | diagnosis MRI only | ceT1w, nT1w, T2w, FLAIR | 66.6 | 35.4 | 22.4 | 20.1 | 29.2 | 21.2 |
| $CNN_{ce}$ | | ceT1w, | 68.6 | 34.2 | 21.9 | 19.3 | 29.0 | 20.0 |
| $CNN_{ce+FLAIR}$ | | ceT1w, FLAIR | 62.6 | 35.2 | 18.9 | 16.4 | 25.8 | 17.5 |

**Abbrev.:** PPV, positive predictive value; ceT1w, contrast-enhanced T1-weighting; FLAIR, fluid-attenuated inversion recovery MRI, SD, standard deviation; TP: time point with TP 0 corresponding to pre-diagnosis and TP 1 corresponding to diagnosis time point.

**Supplemental Digital Content 3:** Results of significance tests.

| **Measurement** | **CNN 1** | **CNN 2** | **Uncorrected pvalue** |
|---|---|---|---|
| FP | $CNN_{dual_ce}$ | $CNN_{dual_all}$ | 3.1e-6 |
| Sensitivity | $CNN_{dual_ce}$ | $CNN_{dual_all}$ | 4.9e-2 |
| PPV | $CNN_{dual_ce}$ | $CNN_{dual_all}$ | 3.1e-4 |
| F1 | $CNN_{dual_ce}$ | $CNN_{dual_all}$ | 1.5e-4 |
| Sensitivity | $CNN_{dual_ce}$ | $CNN_{dual_Flair}$ | 2.9e-9 |
| PPV | $CNN_{dual_ce}$ | $CNN_{dual_Flair}$ | 9.7e-7 |
| F1 | $CNN_{dual_ce}$ | $CNN_{dual_Flair}$ | 1.7e-9 |
| Sensitivity | $CNN_{dual_ce}$ | $CNN_{ce}$ | 3.5e-1 |
| PPV | $CNN_{dual_ce}$ | $CNN_{ce}$ | 1.5e-6 |
| F1 | $CNN_{dual_ce}$ | $CNN_{ce}$ | 9.4e-5 |
| Sensitivity | $CNN_{dual_ce}$ | $CNN_{dual_ce+FLAIR}$ | 4.1e-1 |
| PPV | $CNN_{dual_ce}$ | $CNN_{dual_ce+FLAIR}$ | 7.3e-1 |
| F1 | $CNN_{dual_ce}$ | $CNN_{dual_ce+FLAIR}$ | 5.9e-1 |
| Sensitivity | $CNN_{dual_ce}$ | $CNN_{ce+FLAIR}$ | 9.7e-1 |
| PPV | $CNN_{dual_ce}$ | $CNN_{ce+FLAIR}$ | 1.5e-8 |
| F1 | $CNN_{dual_ce}$ | $CNN_{ce+FLAIR}$ | 1.5e-6 |
| Sensitivity | $CNN_{dual_all}$ | $CNN_{all}$ | 1.0e-2 |
| PPV | $CNN_{dual_all}$ | $CNN_{all}$ | 6.6e-1 |
| F1 | $CNN_{dual_all}$ | $CNN_{all}$ | 4.9e-1 |
| Sensitivity | $CNN_{ce}$ | $CNN_{all}$ | 2.6e-1 |
| PPV | $CNN_{ce}$ | $CNN_{all}$ | 9.7e-1 |
| F1 | $CNN_{ce}$ | $CNN_{all}$ | 9.2e-1 |
| Sensitivity | $CNN_{ce}$ | $CNN_{T1n_ce}$ | 3,9e-2 |
| PPV | $CNN_{ce}$ | $CNN_{T1n_ce}$ | 4,0e-1 |
| F1 | $CNN_{ce}$ | $CNN_{T1n_ce}$ | 9,1e-1 |

Results of all reported significance tests. Bonferroni-corrected significance level is p = 2e-3, threshold for high significance is p = 4e-4. Significant results are greyed.